\documentclass[aps,pra,twocolumn, superscriptaddress]{revtex4}
\usepackage{longtable}
\usepackage{graphicx}
\usepackage{dcolumn}
\usepackage{amsmath}
\usepackage{amssymb}
\usepackage{subfigure}
\usepackage{color,xcolor}
\usepackage{float}
\usepackage{epstopdf}

\def\beq{\begin{equation}}
\def\eeq{\end{equation}}

\begin{document}
\title{Density Functionals that Recognize Covalent, Metallic, and Weak Bonds}

\author{Jianwei Sun}
\affiliation{Department of Physics and Engineering Physics and Quantum Theory Group,
Tulane University, New Orleans, Louisiana 70118, USA}
\author{Bing Xiao}
\affiliation{Department of Physics and Engineering Physics and Quantum Theory Group,
Tulane University, New Orleans, Louisiana 70118, USA} 
\author{Yuan Fang}
\affiliation{Department of Physics and Engineering Physics and Quantum Theory Group,
Tulane University, New Orleans, Louisiana 70118, USA} 
\author{Robin Haunschild}
\affiliation{Department of Chemistry, Rice University, Houston, Texas 77005, USA} 
\author{Pan Hao}
\affiliation{Department of Physics and Engineering Physics and Quantum Theory Group,
Tulane University, New Orleans, Louisiana 70118, USA} 
\author{Adrienn Ruzsinszky}
\affiliation{Department of Physics and Engineering Physics and Quantum Theory Group,
Tulane University, New Orleans, Louisiana 70118, USA} 
\author{G$\acute{{\rm a}}$bor I. Csonka}
\affiliation{Department of Inorganic and Analytical Chemistry, Budapest University of Technology and Economics, H-1521 Budapest, HUNGARY} 
\author{Gustavo E. Scuseria}
\affiliation{Department of Chemistry, Rice University, Houston, Texas 77005, USA}
\affiliation{Department of Physics and Astronomy, Rice University, Houston, Texas 77005, USA}
\affiliation{Chemistry Department, Faculty of Science, King Abdulaziz University, Jeddah 21589, Saudi Arabia}
\author{John P. Perdew}
\affiliation{Department of Physics and Engineering Physics and Quantum Theory Group, Tulane University, New Orleans, Louisiana 70118, USA} 

\date{\today}

\begin{abstract}
Computationally-efficient semilocal approximations of density functional theory at the level of the local spin density approximation (LSDA) or generalized gradient approximation (GGA) poorly describe weak interactions. We show improved descriptions for weak bonds (without loss of accuracy for strong ones) from a newly-developed semilocal meta-GGA (MGGA), by applying it to molecules, surfaces, and solids. We argue that this improvement comes from using the right MGGA dimensionless ingredient to recognize all types of orbital overlap.

\


\end{abstract}

\maketitle

Due to its computational efficiency and reasonable accuracy, the Kohn-Sham density functional theory \cite{KS, Parr_Yang, PK} with semilocal approximations  to the exchange-correlation energy, e.g., the local spin density approximation (LSDA)~\cite{PW92, SPS_PRB_2010} and the standard Perdew-Burke-Ernzerhof  (PBE) generalized gradient approximation (GGA)~\cite{PBE}, is one of the most widely-used electronic structure methods in materials science, surface science, condensed matter physics, and chemistry. Semilocal approximations display a well-understood error cancellation between exchange and correlation in bonding regions. Thus some intermediate-range correlation effects, important for strong and weak bonds, are carried by the {\it exchange} part of the approximation. However, it is well-known that these approximations cannot yield correct long-range asymptotic dispersion forces~\cite{JG_RMP_1989}. This raises doubts about the suitability of semilocal approximations for the description of weak interactions (including hydrogen bonds and van der Waals interactions), even near equilibrium where most interesting properties occur. These doubts are supported by the performance of LSDA and GGAs, which are not very useful for many important systems and properties (such as DNA, physisorption on surfaces, most biochemistry, etc.).  

However, these doubts are challenged by recent developments in semilocal meta-GGAs (MGGA)~\cite{TPSS, ZT_JCP_2006, PRCCS, SXR_JCP_2012, CGTV_CPL_2012, HSXRCTP_JCTC_2012, SHXSP} (which are useful by themselves and as ingredients of hybrid functionals~\cite{SHXSP}). Compared to GGAs, which use the density $n({\bf r})$ and its gradient $\nabla n$ as inputs, MGGAs additionally include the positive kinetic energy density $\tau=\sum_{k} \left |\nabla \psi_{k} \right |^2/2$ of the occupied orbitals $\psi_{k}$. For simplicity, we suppress the spin here. By including training sets of noncovalent interactions, the molecule-oriented and heavily-parameterized M06L MGGA was trained to capture medium-range exchange and correlation energies that dominate equilibrium structures of noncovalent complexes \cite{ZT_JCP_2006}. Madsen {\it et al.} showed that the inclusion of the kinetic energy densities enables MGGAs to discriminate between dispersive and covalent interactions, which makes the M06L MGGA~\cite{ZT_JCP_2006} suitable for layered materials bonded by van der Waals interactions~\cite{MFH_JPCL_2010, AHH_PRB_2012}. Besides improvement for noncovalent bonds, simultaneous improvement for metallic and covalent bonds is also an outstanding problem for semilocal functionals\cite{SHSGMMK, SMRKP}. Ref.~\onlinecite{SMRKP} has shown that the revised Tao-Perdew-Staroverov-Scuseria (revTPSS)~\cite{PRCCS} MGGA, due to the inclusion of the kinetic energy density, simultaneously predicts accurate results for the adsorption energy of CO on the Pt (111) surface and the lattice constant and surface energy of the substrate, while GGA and LSDA \cite{SMRKP} do not. These successful applications show the advantages and flexibility brought by inputting the kinetic energy density for MGGAs. The effect of the dependence on the kinetic energy density has also been studied recently in Refs.~\onlinecite{ SXR_JCP_2012, SHXSP}, leading to a new MGGA called MGGA\_MS2~\cite{SHXSP}. However, an important question for the development of MGGAs and thus for understanding the results from MGGAs remains unanswered: how should the kinetic energy density be built into MGGAs, why, and what are the consequences?

To answer this question, let's look at the input, the positive kinetic energy density, and how it is used in the M06L, revTPSS, and MGGA\_MS2 MGGAs. $\tau$ can be expressed explicitly in terms of the density $n({\bf r})$ for two kinds of paradigm systems. For one- and two-electron ground states, 
$\tau$ reduces to the von Weizs$\ddot{{\rm a}}$cker kinetic energy density, $\tau^W=\frac{1}{8}|\nabla n|^2/n$. These systems, especially the hydrogen atom, are paradigmatic for quantum chemistry. On the other hand, for a uniform electron gas of density $n$, which is a paradigm density for solids and a constraining limit for most approximate functionals, $\tau$ becomes $\tau^{\rm unif}=\frac{3}{10}(3 \pi^2)^{2/3}n^{5/3}$. In the uniform electron gas, electrons are fully delocalized and orbitals are highly overlapped. 

\begin{center}
\noindent
\begin{table}[htb]
\caption{Values of $z$, $\alpha$, and $t^{-1}$ for typical regions.}
\centering
\vspace{2mm}
\begin{tabular}{l c c c}
\hline\hline

	Region& $z$	&$\alpha$&$t^{-1}$\\
\hline						
Single orbital	&1	&0&5$s^2$/3\\
Slowly-varying density	&$\approx 0$	&$\approx 1$&$\approx 1$\\
Overlap of closed shells	&$\approx 0$	&$\gg 1$&$\gg 1$\\
\hline
\end{tabular}

\label{table:MGGA_variable}
\end{table}
\end{center}

With $\tau^W$ and $\tau^{\rm unif}$, three different dimensionless parameters can be constructed: $z=\tau^W/\tau$, $\alpha=(\tau-\tau^W)/\tau^{\rm unif}$, and $t^{-1}=\tau/\tau^{\rm unif}=\alpha+5s^2/3$. Here, $s=|\nabla n|/[2(3 \pi ^2)^{1/3} n ^{4/3}]$ is the reduced density gradient, an important dimensionless parameter measuring the inhomogeneity of the density and employed in GGAs. Table~\ref{table:MGGA_variable} shows the values of these three parameters for three different typical regions: 1) regions of one- and two-electron densities that characterize single bonds; 2) regions of slowly-varying density that characterize metallic bonds; and 3) regions of density overlap between closed shells that characterize noncovalent bonds. It can be seen that only $\alpha$ can recognize all types of orbital overlap and should be chosen as the ingredient for constructing MGGAs. It equals 0 for the regions of one- and two-electron densities and $\approx 1$ for regions of slowly-varying density. For regions of density overlap between closed shells, $\tau^W \approx 0$ because the density gradient is small by symmetry, $\tau/\tau^{\rm unif} \propto n/n^{5/3}$, and thus $\alpha \gg 1$ as $n$ is small in such regions. Note $\alpha \ge 0$ since $\tau^W$ is a lower bound on $\tau$~\cite{KPB_IJQC_1999}.

However, instead of $\alpha$, which is only used for recovering the fourth-order gradient expansion of the exchange energy of a slowly-varying density, revTPSS uses $z$ to identify different orbital-overlap regions. $z$ identifies the single-orbital region with $z=1$ and the slowly-varying density with $z \approx 0$, but it cannot distinguish regions of slowly-varying density from those of density overlap between closed shells, both of which have $z \approx 0$. Since revTPSS used the information of the hydrogen atom (where $z=1$) and the slowly-varying density (where $z \approx 0$) to fix its parameters and orient toward covalent bonds, this limitation of $z$ makes revTPSS unable to track noncovalent bonds. This is manifested by its performance on the problem of graphene adsorbed on the Ni(111) surface. 

\begin{figure*}
{\subfigure[]{\label{figure:Graphene-Ni}\includegraphics[width=0.41\linewidth]{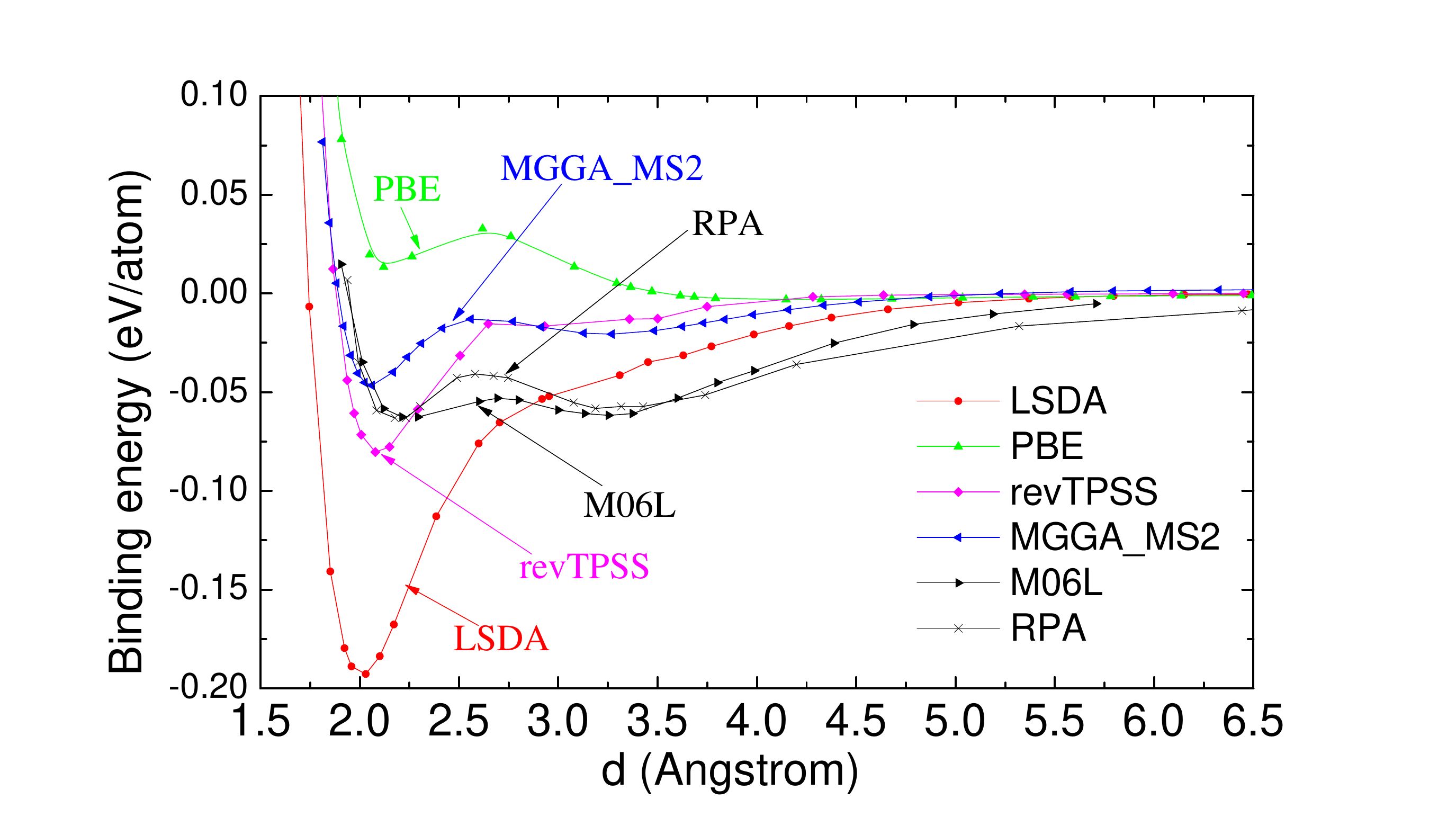}}}
{\subfigure[]{\label{figure:Ar2_rs_s_alpha}\includegraphics[width=0.41\linewidth]{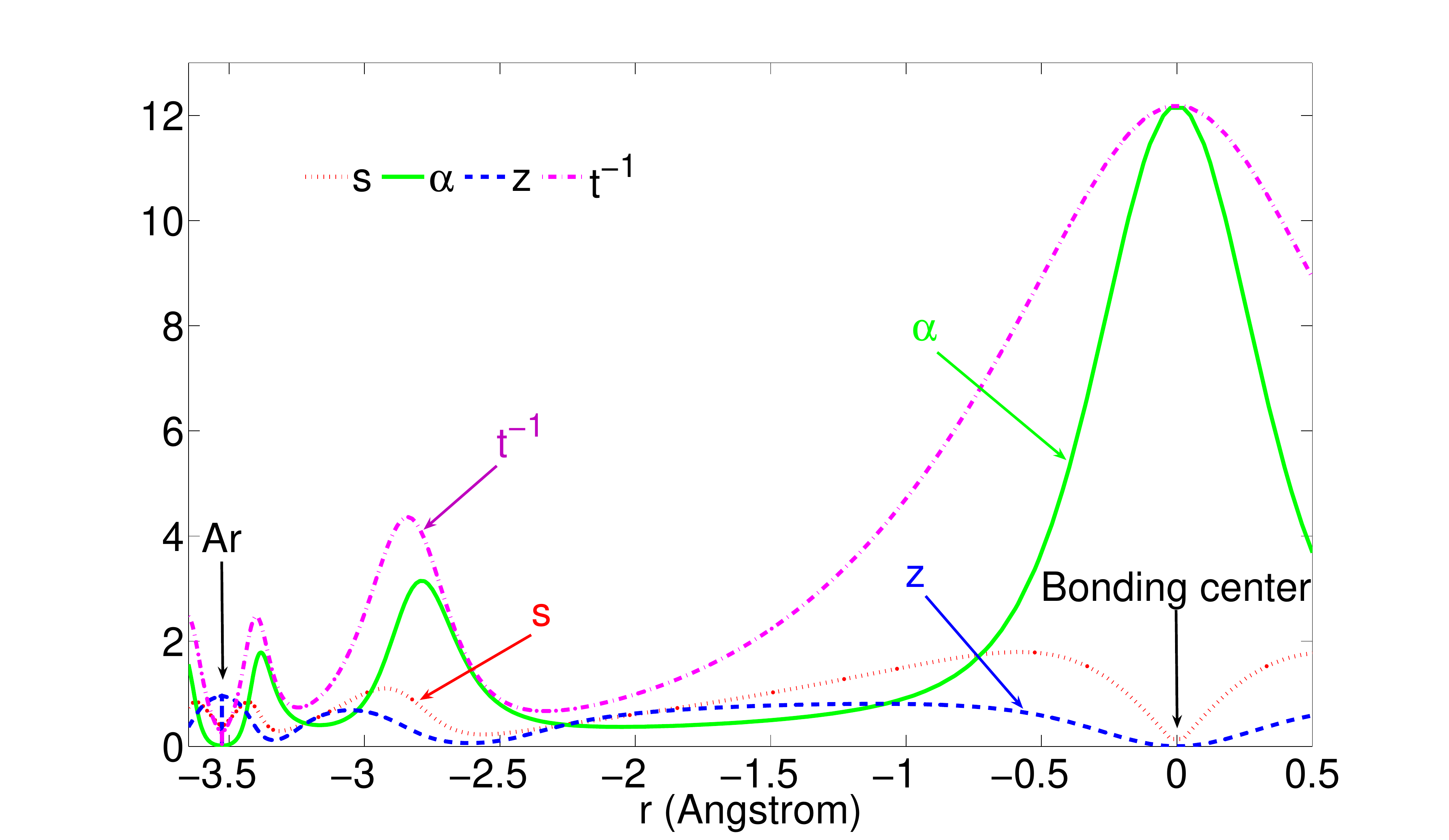}}}
{\subfigure[]{\label{figure:Fx_s_alpha}\includegraphics[width=0.41\linewidth]{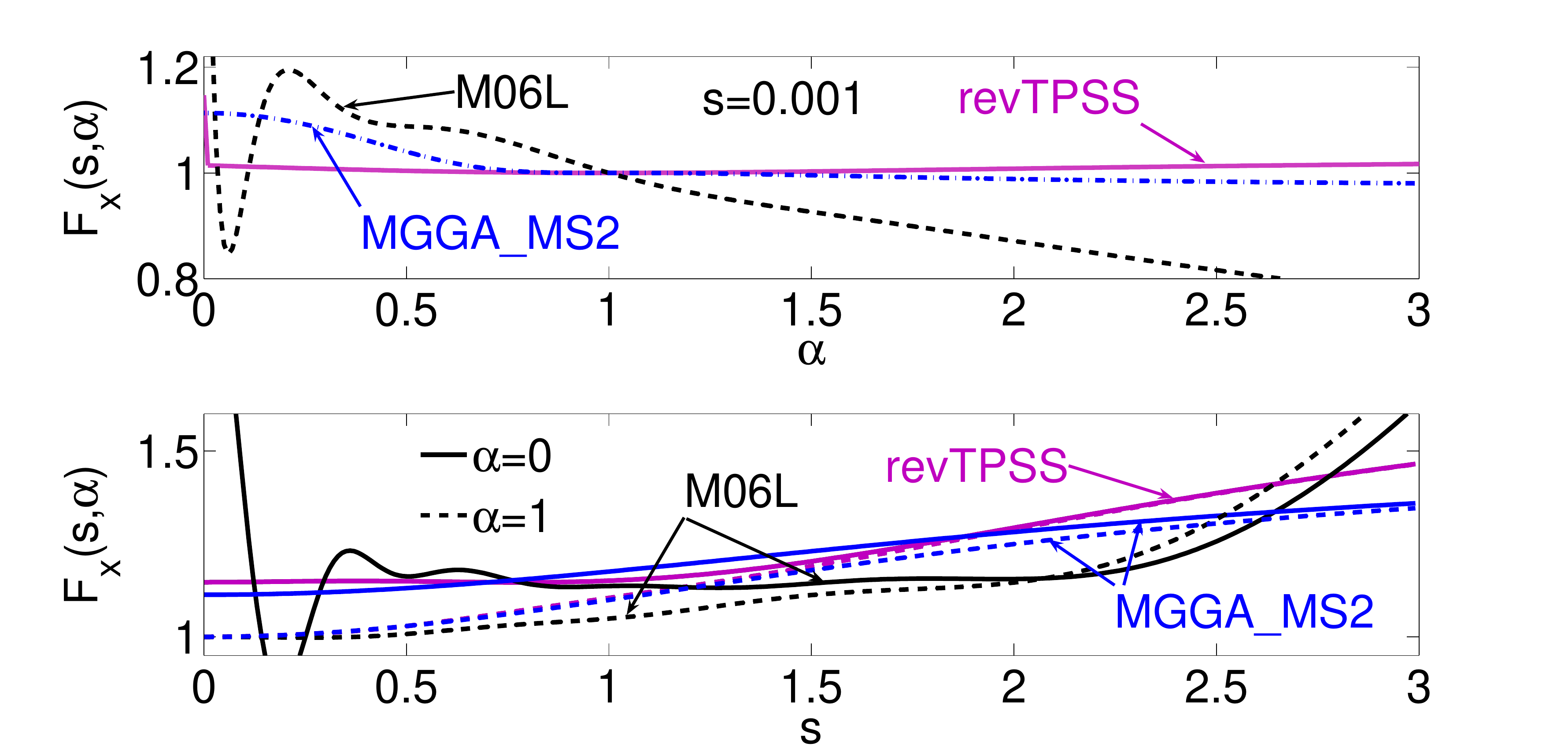}}}
{\subfigure[]{\label{figure:Ar2_BC}\includegraphics[width=0.41\linewidth]{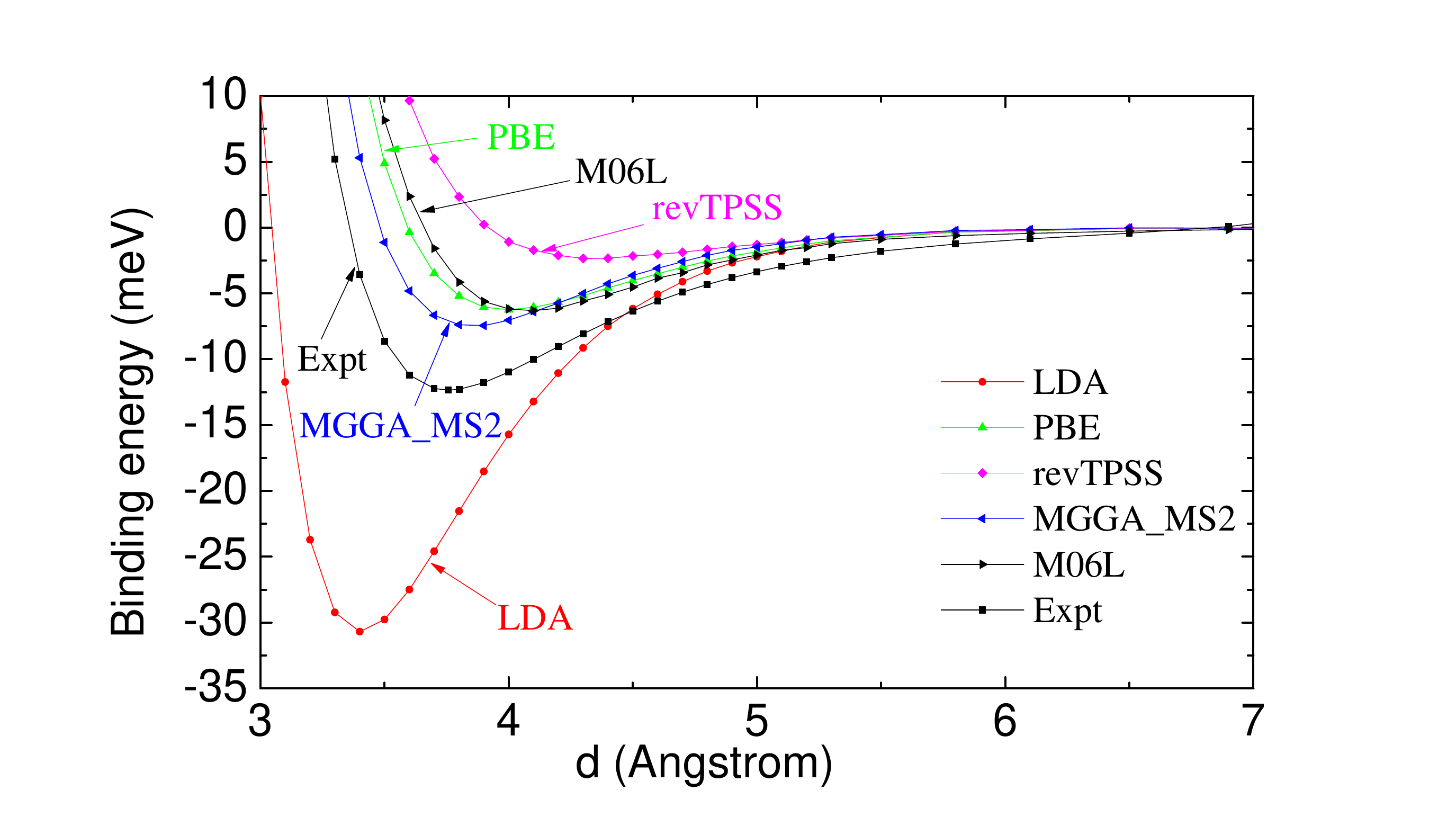}}}
\caption{(a) Binding-energy curves for graphene adsorption on the Ni (111) surface. The M06L and RPA values are from Refs.~\onlinecite{AHH_PRB_2012} and~\onlinecite{MGRKHK_PRB_2011}, respectively. b) Distributions of $s$, $\alpha$, $z$, and $t^{-1}$ for Ar$_2$ at equilibrium. (c) Exchange enhancement factor. (d) Binding-energy curve of Ar$_2$ for different functionals.  See Ref.~\onlinecite{Suppl} for computational details.}
\label{figure:Ar2}
\end{figure*}

Graphene was found to bind to the Ni (111) surface at 2.11$\pm$0.07 \mbox{\AA} chemically (from experiment~\cite{GNWTO_SS_1997}) and at 3.3 \mbox{\AA} physically (from theory~\cite{MGRKHK_PRB_2011}), respectively. In Fig.~\ref{figure:Graphene-Ni}, the binding curve of the random phase approximation (RPA) from Ref~\onlinecite{MGRKHK_PRB_2011}, which naturally includes the nonlocal van der Waals interaction, is included as the reference. RPA predicts both the chemisorption and the physisorption minima with the correct long-range asymptotic behavior~\cite{MGRKHK_PRB_2011}. Note that RPA calculations are  about 2 orders of magnitude more time consuming than those of semilocal functionals. Fig.~\ref{figure:Graphene-Ni} shows that LSDA predicts the chemisorption minimum but misses the physisorption one, while PBE misses the latter and almost misses the former. Compared to the RPA results, revTPSS predicts the chemisorption with a better binding energy than those of LSDA and PBE. But it still misses the physisorption minimum, consistent with the result of Ref.~\onlinecite{WLMPLNBJ_PRB_2012}, indicating its inability to capture noncovalent bonds. Although all semilocal functionals considered here yield wrong long-range asymptotic behaviors, MGGA\_MS2 and M06L remarkably capture the double minima at accurate distances, demonstrating their ability to describe both strong and weak bonds. If we use the experimental and RPA binding distances (2.11$\pm$0.07 \mbox{\AA}, 3.3 \mbox{\AA}) as references for the chemisorption and physisorption, respectively, then MGGA\_MS2 (2.09 \mbox{\AA}, 3.29 \mbox{\AA}) yields a better agreement than M06L (2.29 \mbox{\AA}, 3.25 \mbox{\AA})~\cite{AHH_PRB_2012}. In terms of the binding energies, M06L (64 meV, 64 meV)~\cite{AHH_PRB_2012} is better than MGGA\_MS2 (49 meV, 23 meV) when compared to RPA (67 meV, 60 meV)~\cite{MGRKHK_PRB_2011}, with the first number in the parentheses for the chemisorption and the second for the physisorption.


In MGGA\_MS2~\cite{SHXSP}, $\alpha$ is the only ingredient built from $\tau$, and it can recognize all types of orbital overlap. Although the two fitting parameters of MGGA\_MS2 are determined by two data sets that do not involve weak interactions~\cite{SHXSP}, the ability of $\alpha$ to identify the overlap between closed shells helps MGGA\_MS2 to capture weak interactions near equilibrium, which explains why MGGA\_MS2 captures both minima in Fig.~\ref{figure:Graphene-Ni}.

M06L, on the other hand, uses  $t^{-1}$ to identify different orbital-overlap, with $\alpha$ only to make the one-electron correlation energy exact. Table~\ref{table:MGGA_variable} shows that $t^{-1} \approx 1$ for the slowly-varying density, and $t^{-1} \gg 1$ for the overlap between two closed shells. Madsen {\it et al.} have shown that $t^{-1}$ can discriminate between covalent and noncovalent interactions~\cite{MFH_JPCL_2010}, which then explains the ability of M06L to capture the double minima in Fig.~\ref{figure:Graphene-Ni}. This example together with many other successful applications of M06L, including CO adsorbed on the Pt(111) surface~\cite{LZT_JPCL_2012} and layered solids~\cite{MFH_JPCL_2010}, seems to suggest that $t^{-1}$, in addition to $s$, would be a good dimensionless parameter that measures the inhomogeneity of the density. 
However, $t^{-1}$ loses the ability to identify the single-orbital regions, where $t^{-1}=5s^2/3$ provides no extra information in comparison with $s$. It can be seen in Fig.~\ref{figure:Ar2_rs_s_alpha} that, for the region around the nucleus of Ar where the {\it 1s} orbital dominates, $t^{-1}$ resembles $s$ in shape. To overcome the shortcoming of $t^{-1}$ being unable to identify single-orbital regions, and to fit to a large number of data sets that include covalent single bonds ($\alpha \approx 0$), 35 fitting parameters had to be introduced into M06L, causing an oscillation in the exchange enhancement factor near small $\alpha$ as shown in Fig.~\ref{figure:Fx_s_alpha}. The exchange enhancement factor {\it F$_x$} characterizes the enhancement of the exchange energy density with respect to its local approximation, defined by $E_x[n]=\int d^3 r n \epsilon_x^{\rm unif}(n)F_x(s, \alpha)$, where $\epsilon_x^{\rm unif}(n)=-3(3 \pi^2 n)^{1/3}/{4\pi}$ is the exchange energy per electron of the uniform electron gas. 

The oscillation in the M06L {\it F$_x$} can cause two major problems. The first one is the lack of computational stability and the high requirements to converge calculations. An example is the oscillation of the binding curve of the Ar dimer shown in Ref~\onlinecite{Suppl}. The oscillation can be removed by using much higher computational settings for the calculations, but even then M06L still yields a too-shallow minimum far away from the experimental binding distance, as shown in Fig.~\ref{figure:Ar2_BC}. The performance of M06L on the Ar dimer and its inability to bind the Kr$_2$ and Xe$_2$ rare gas dimers~\cite{GG11}, which are typical simple tests for van der Waals interactions, contradict its good performance on many other noncovalent systems. This also leads to the second problem, the consistency in describing different bonds in different chemical environments. To investigate this, we study the performance of M06L on systems that are far outside its training sets, e.g., lattice constants of solids. We choose the six ionic insulators and semiconductors studied in Ref~\onlinecite{ZTPAS_PRL_2011}. Ref~\onlinecite{ZTPAS_PRL_2011} showed that, although the covalent and ionic bonds dominate, the van der Waals interaction plays an essential role in determining accurate lattice constants. It found that the errors in lattice constants of these six solids are reduced by a factor of 2, in comparison with experimental data, when the van der Waals interactions were included on top of the PBE functional. 

Table~\ref{table:Latt_ionic_semi} shows that LDA understimates the lattice constants of the 6 solids while PBE overestimates them. revTPSS improves the lattice constants over PBE slightly for ionic insulators but significantly for semiconductors when compared to experimental data. However, revTPSS still yields too large lattice constants because revTPSS misses weak interactions in these solids. Therefore, the improvement of revTPSS over PBE should be ascribed to its better description on covalent single bonds, by which those semiconductors are bound. Adding the van der Waals interactions on top of revTPSS would be expected to further improve the lattice constants. This is largely realized by MGGA\_MS2 alone, which reduces the mean absolute error (MAE) from 0.042 \mbox{\AA} of revTPSS to a remarkable 0.011 \mbox{\AA}. This demonstrates one more time that MGGA\_MS2 describes well both strong (covalent and ionic) bonds and weak interactions because $\alpha$ can recognize all types of orbital overlap.

The pattern of the M06L performance on these 6 solids however is inconsistent. Compared to experimental data, M06L predicts accurate lattice constants for C and Si. But it yields too-large ones, even larger than those of PBE, for Ge and GaAs which have relatively large ionic radii and thus significant van der Waals interactions. Similar behavior is also observed for the two ionic solids. M06L yields an accurate lattice constant for MgO, but a too-large one for NaCl. This random performance is likely related to the oscillation of the exchange enhancement factor near small $\alpha$ and small $s$ as shown in Fig.~\ref{figure:Fx_s_alpha}, because small $s$ is more relevant to solids than to molecules. The inconsistent performance of M06L in this test set cautions against its use in solids, and is rooted in using $t^{-1}$ to construct the functional. 

\begin{center}
\noindent
\begin{table}[htb]
\caption{Errors in lattice constants of ionic insulators and semiconductors from different functionals. The zero-point anharmonic expansion has been removed from the experimental lattice constants~\cite{HFSCPP_PRB_2012}. ME: mean error; MAE: mean absolute error.}
\centering
\vspace{2mm}
\begin{tabular}{l c c c c c c}
\hline\hline

solids	&LDA	&PBE	&M06L	&revTPSS	&MGGA\_MS2	&Expt. \\
\hline	
C&	-0.022&	0.014	&-0.010	&0.003	&-0.007&	3.555\\
Si&	-0.017&	0.046&	-0.010	&0.017	&0.005&5.422\\
Ge	&-0.013&	0.124&	0.137	&0.038	&0.005&	5.644\\
GaAs	&-0.026	&0.111	&0.143	&0.039	&0.002	&5.641\\
NaCl	&-0.098	&0.130	&0.117	&0.102	&0.029	&5.565\\
MgO	&-0.018	&0.073&0.012	&0.052	&0.019	&4.188\\
\hline	
ME	&-0.032	&0.083&	0.065	&0.042	&0.009	\\
MAE&	0.032	&0.083	&0.072	&0.042	&0.011	\\

\hline
\end{tabular}

\label{table:Latt_ionic_semi}
\end{table}
\end{center}

The above discussions qualitatively explain the performance of revTPSS, M06L, and MGGA\_MS2, with a special focus on noncovalent interactions, whereas quantitative analysis using enhancement factors is a subtle issue. Madsen {\it et al.}~\cite{MFH_JPCL_2010} argued that the inability of TPSS~\cite{TPSS} (whence also of revTPSS) to capture weak interactions is due to the increase of its {\it F$_x$} at small $s$ as $\alpha$ becomes large, as shown in Fig.~\ref{figure:Fx_s_alpha} for revTPSS. That argument is supported by the fact that the MGGA\_MS2 {\it F$_x$} at small $s$ {\it decreases monotonically} as $\alpha$ becomes large; see Fig.~\ref{figure:Fx_s_alpha} of this work and Fig. 2 of Ref.~\onlinecite{SHXSP}. The monotonically decreasing $\alpha$-dependence can be rationalized by the fact that the strength of different chemical bonds decreases with increasing $\alpha$ (e.g., $\alpha=0$ for single bonds, $\alpha \approx 1$ for metallic bonds, and $\alpha \gg 1$ for noncovalent bonds.) Note that $\alpha$ is related to the electron localization function~\cite{BE_JCP_1990}, $\eta$=1/(1+$\alpha^2$), which has been used to establish a rigorous topological classification of chemical bonds~\cite{SS_Nature_1994}, but surprisingly was not recognized as a useful ingredient for MGGAs. However, it should be stressed that the improvement for weak interactions near equilibrium from MGGA\_MS2 results from the good balance between its $s-$ and $\alpha-$dependences~\cite{SHXSP, Suppl}.

\begin{figure}
{\includegraphics[width=0.48\linewidth]{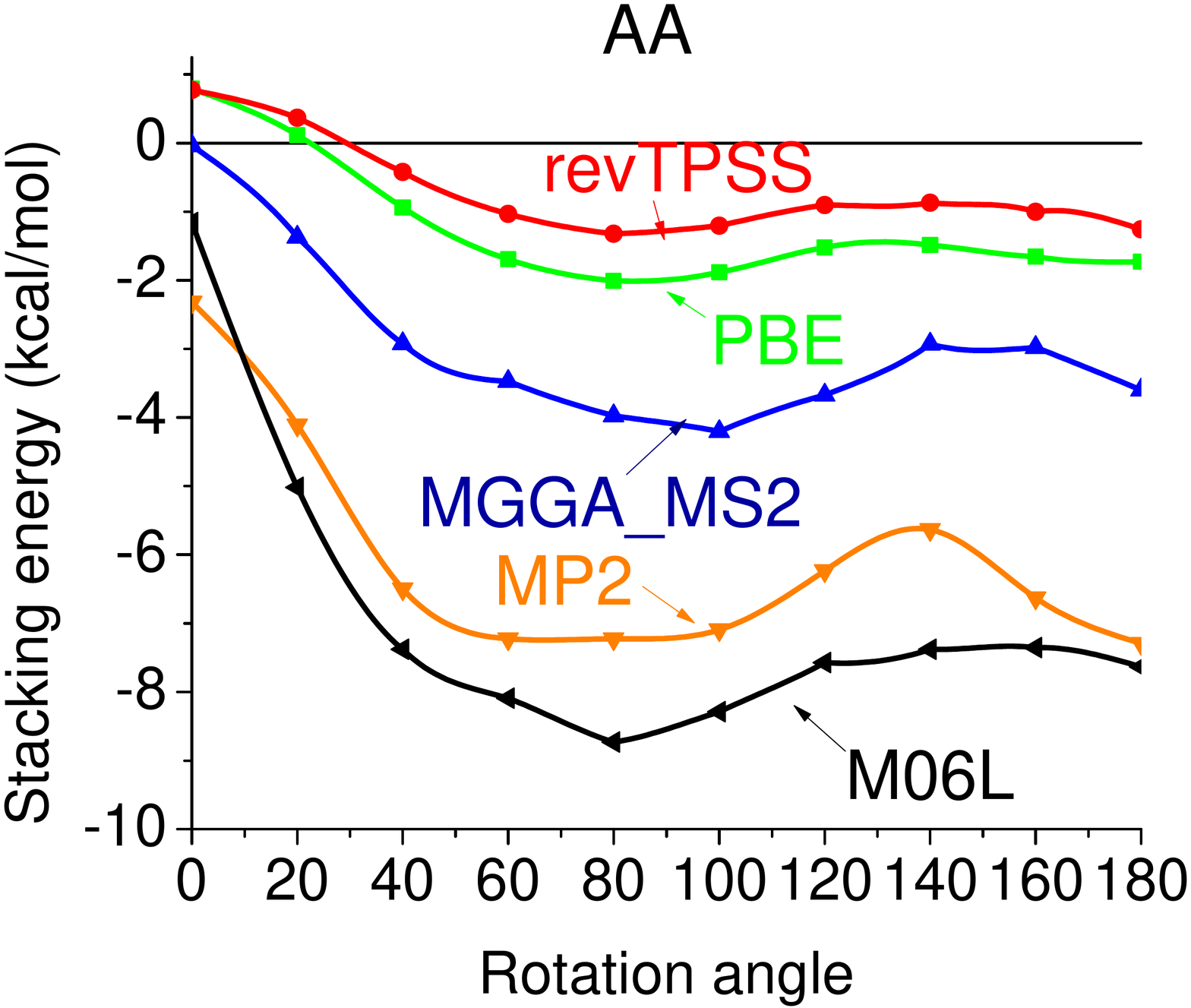}}
{\includegraphics[width=0.48\linewidth]{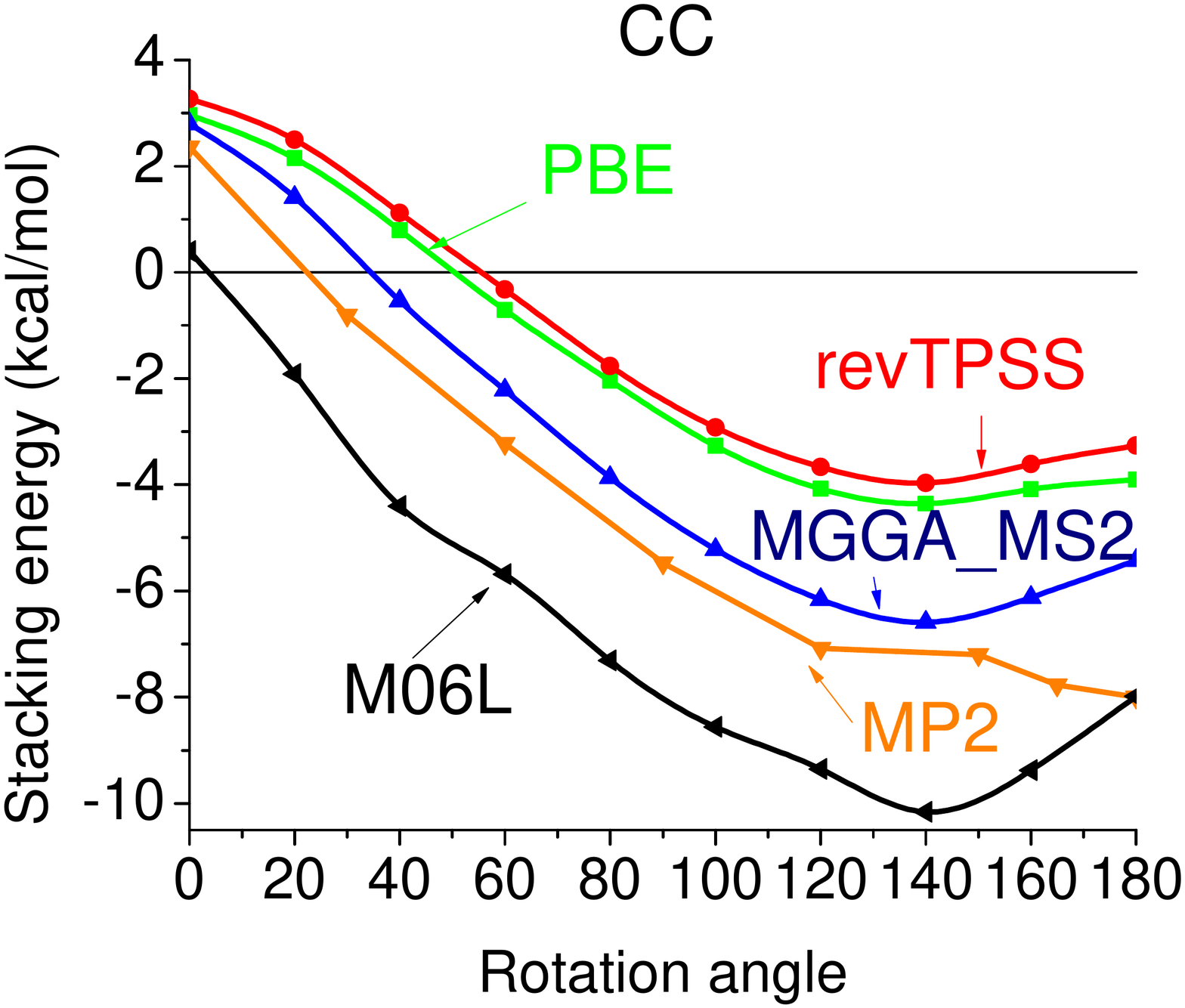}}
{\includegraphics[width=0.48\linewidth]{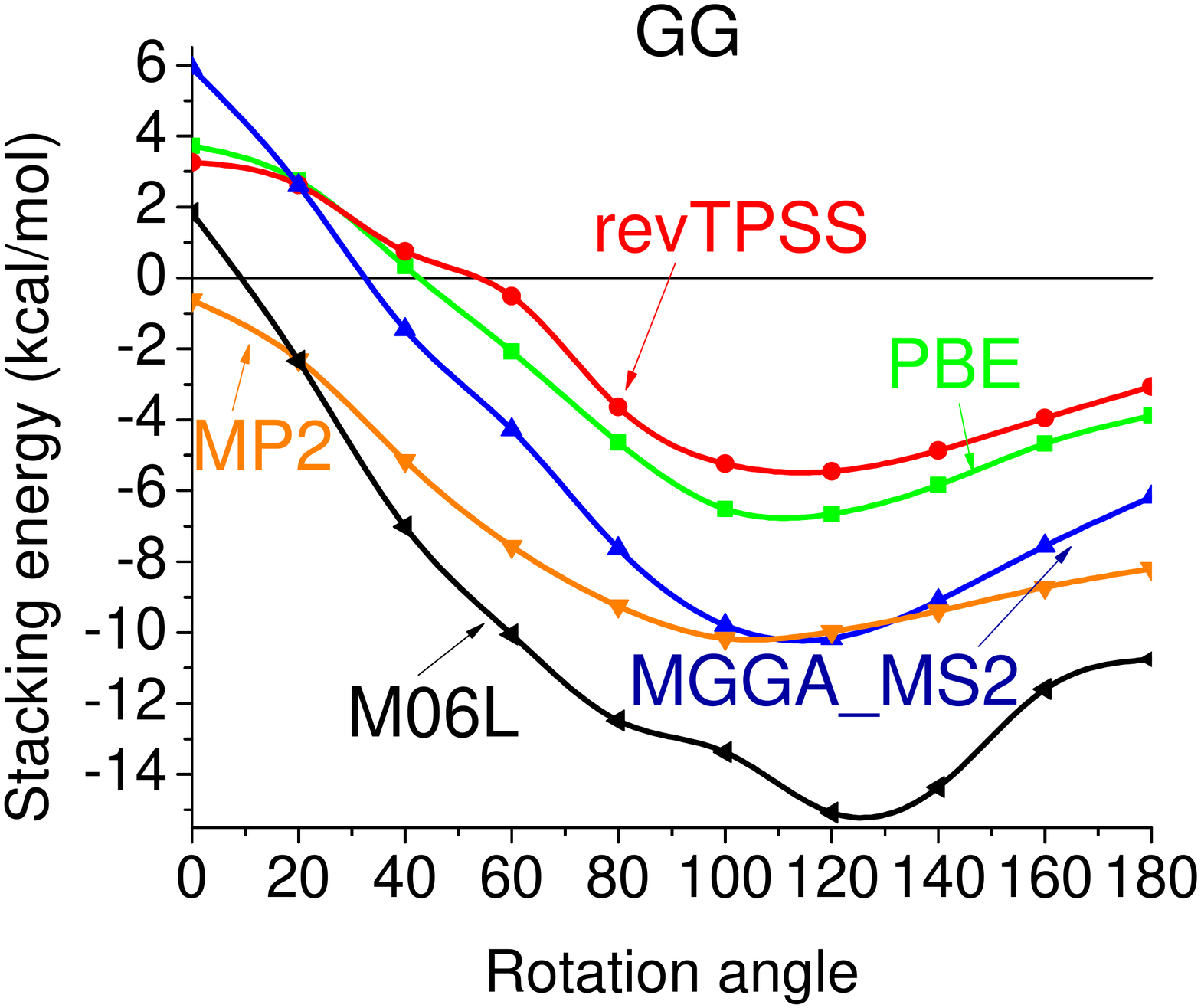}}
{\includegraphics[width=0.48\linewidth]{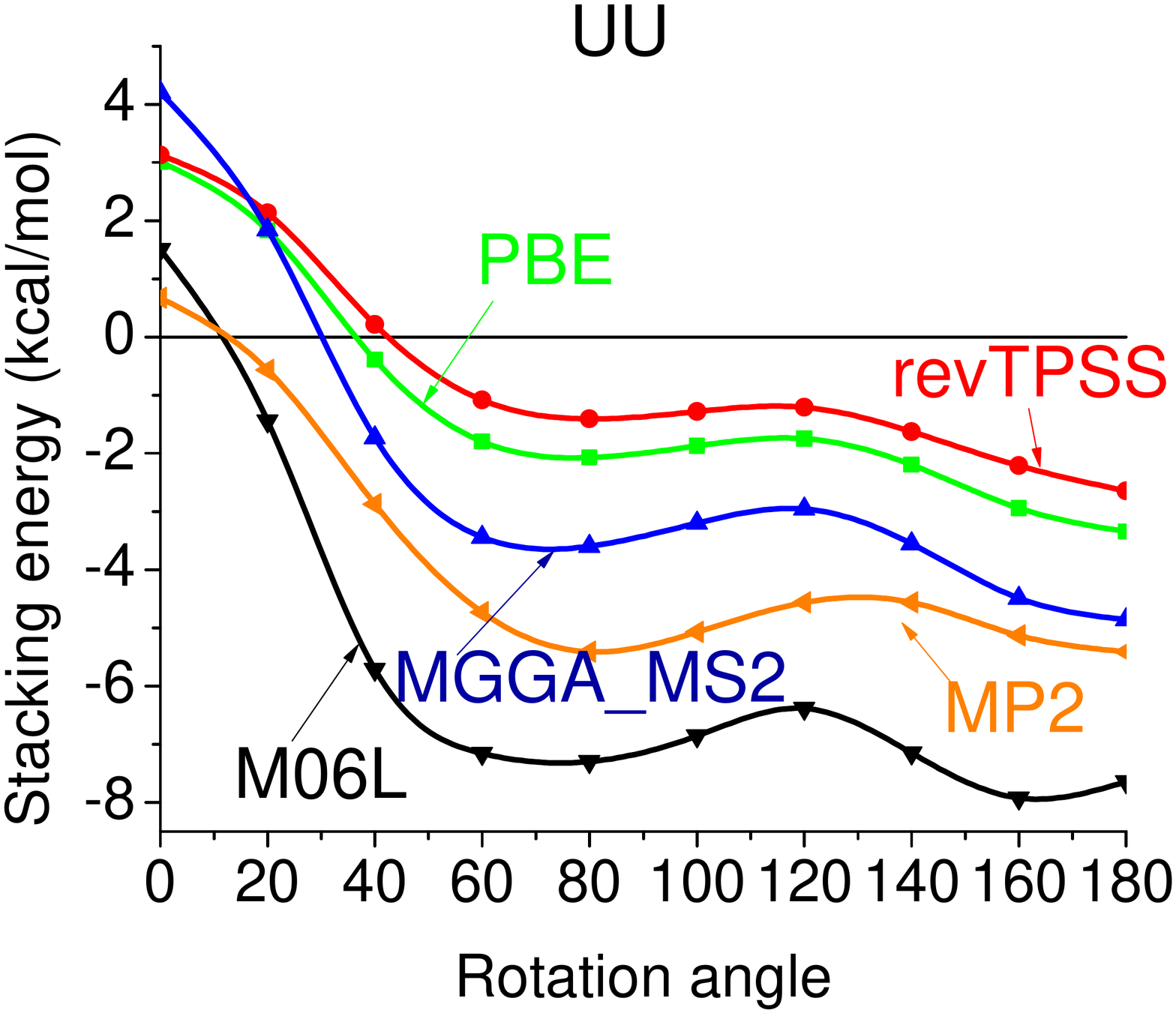}}
\caption{Interaction energy (in kcal/mol) as a function of rotation angle (in degrees) for stacked homonucleobase dimers. The MP2 values were calculated at a fixed separation distance of 3.4 \AA, taken from Ref.~\onlinecite{CTL_JCP_2008}. The PBE, revTPSS, M06L,  and MGGA\_MS2 values were calculated at their own optimized distances for 180$^\circ$. See Ref.~\onlinecite{Suppl} for computational details. 1 kcal/mol = 0.0434 eV.}
\label{figure:base}
\end{figure}

To show its potential impact on a wide range of research and also to test its consistency for different systems, we apply MGGA\_MS2 for stacking energies of nucleobases, which are essential for the structural stability and function of DNA and RNA~\cite{CTL_JCP_2008, YPF_NAR_2006}. 
Fig.~\ref{figure:base} shows how the interaction energies for several stacked pairs of nucleobases (from adenine (A), cytosine (C), guanine (G), and uracil (U)) vary as one nucleobase rotates relative to the other around their perpendicular axis. The reference point for the rotation was defined using the method of Elstner {\it et al.}~\cite{EHFSK_JCP_2001}, in which the center of mass of each nucleobase was aligned with the glycocidic bonds parallel. A counterclockwise (right-handed) rotation was then applied (see Fig. 2 of Ref.~\onlinecite{CTL_JCP_2008}). In each case, the nucleobase structure was fixed to that of the optimized isolated molecule.  Except for the CC stacking structures, where MGGA\_MS2 misses the minimum at 180$^\circ$ of MP2 (M$\o$ller-Plesset second-order perturbation theory in the electron-electron interaction), MGGA\_MS2 yields the stacking energies qualitatively in agreement with the reference MP2 values~\cite{CTL_JCP_2008}. For all considered cases where binding between nucleobases is expected, MGGA\_MS2 significantly improves the stacking energies over PBE and revTPSS, but still underestimates them about the same amount as M06L overestimates, in comparison with the reference MP2 method. 
If this performance persists for other pairs, then MGGA\_MS2 can provide a better description for DNA and RNA conformations than PBE (and, when long-range van der Waals interactions are added~\cite{GG11}, probably better than M06L). This could be an important step toward computer-assisted drug design since the computational cost of the semilocal MGGA\_MS2 is affordable for large bio-molecules.

The computationally-efficient semilocal density functionals must fail for many stretched bonds~\cite{RPCVS_JCP_2006}. But we have suggested that some MGGAs can be usefully accurate for unstretched or modestly-stretched ones, not just for strong but even for weak bonds. Successful but simple MGGAs employ only $\alpha=(\tau-\tau^W)/\tau^{\rm unif}$ to recognize all types of orbital overlap, and extrapolate monotonically from $0 \le \alpha \lesssim 1$ to large $\alpha$. This insight should guide the construction of further-improved semi-empirical and nonempirical density functionals.

\


This work was supported by NSF under
Grant Nos. DMR08-54769 and CHE-1110884, by NSF
Cooperative Agreement No. EPS-1003897 with further
support from the Louisiana Board of Regents, by the
Welch Foundation (C-0036), and by the Deutsche
Forschungsgemeinschaft (HA 5711/2-1). JS thanks Jianmin Tao for helpful discussions. The computations
were made with the support of the Louisiana Optical Network
and the Tulane Center for Computational Sciences.


\clearpage
\newpage

\end{document}